\documentstyle[twocolumn,floats,aps,prb,epsf]{revtex}
\draft

\title{\boldmath Susceptibility Inhomogeneity and Non-Fermi-Liquid Behavior in 
Ce(Ru$_{0.5}$Rh$_{0.5}$)$_2$Si$_2$}
\author{D.~E. MacLaughlin}
\address{Department of Physics, University of California, 
Riverside, California 92521-0413}
\author{O. O. Bernal}
\address{Department of Physics and Astronomy, California State University, Los Angeles, 
California 90032}
\author{J.~E. Sonier}
\address{Department of Physics, Simon Fraser University, Burnaby, Canada V5A 1S6}
\author{R.~H. Heffner}
\address{MS K764, Los Alamos National Laboratory, Los Alamos, New Mexico 87545}
\author{T. Taniguchi, Y. Miyako}
\address{Graduate School of Science, Osaka University, Toyonaka, Osaka 560-0043,Japan
\\ \rm\small version date \today
}

\address{\parbox{137mm}{\bigskip\rm\small
Magnetic susceptibility and muon spin rotation ($\mu$SR) experiments have been carried out to study the effect of structural disorder on the non-Fermi-liquid (NFL) behavior of the heavy-fermion alloy~Ce(Ru$_{0.5}$Rh$_{0.5}$)$_2$Si$_2$. Analysis of the bulk susceptibility in the framework of disorder-driven Griffiths-phase and Kondo-disorder models for NFL behavior yields relatively narrow distributions of characteristic spin fluctuation energies, in agreement with $\mu$SR linewidths that give the inhomogeneous spread in susceptibility. $\mu$SR and NMR data both indicate that disorder explains the ``nearly NFL'' behavior observed above $\sim$2~K, but does not dominate the NFL physics found at low temperatures and low magnetic fields.
\\[6pt] PACS numbers: 71.27.+a, 75.30.Mb, 76.60.Cq.}}
\begin{document} \maketitle
\narrowtext

\section{Introduction} \label{sect:Intro}
The discovery of non-Fermi-liquid (NFL) phenomena in strongly-correlated electron metals raises fundamental questions about the elementary excitations of these systems.\cite{ITP96,VNvS01} The interest in NFL behavior is in large part due to the expected robustness of Landau's Fermi-liquid theory,\cite{VNvS01,Land57} according to which interactions between electrons that do not precipitate a phase transition should not change the Fermi-liquid nature of the low-lying excitations. Many {\em f\/}-electron heavy-fermion alloys are NFL metals.\cite{ITP96} Attempts to explain NFL behavior often invoke the notion of a quantum critical point (QCP) at zero temperature, the critical behavior of which extends to nonzero temperatures and modifies the thermodynamic and transport properties of the metal. 

Recent nuclear magnetic resonance (NMR) and muon spin rotation ($\mu$SR) investigations of NFL alloys have yielded unambiguous evidence that in some of these materials disorder is a major factor in NFL behavior.\cite{BMLA95,MacL00} In such cases NMR and $\mu$SR spectra reflect broad distributions of the local magnetic susceptibility~$\chi({\bf r},T)$, the high-susceptibility end of which arises from regions of the sample that do not exhibit Fermi-liquid paramagnetism. It is clearly of interest to determine susceptibility distributions in a considerable number of NFL systems, in order to understand better the systematic interplay between quantum criticality and disorder in NFL behavior.

\subsection{\boldmath C\lowercase{e}(R\lowercase{u}$_{1-\lowercase{x}}$R\lowercase{h}$_{\lowercase{x}}$)$_2$S\lowercase{i}$_2$}
The Ce(Ru$_{1-x}$Rh$_x$)$_2$Si$_2$ alloy system exhibits a number of magnetic and nonmagnetic ground states as a result of strong electron correlation.\cite{KKFM95} The end compound CeRu$_2$Si$_2$ is a Fermi-liquid heavy-fermion compound with no evidence for magnetic ordering, whereas CeRh$_2$Si$_2$ undergoes an antiferromagnetic (AFM) transition at a N\'eel temperature~$T_N = 35$~K to a state of local-moment AFM order. With decreasing $x$ $T_N$ is suppressed, and vanishes for $x \approx 0.55$. A second region of AFM order in the phase diagram appears for $0.05 \lesssim x \lesssim 0.25$; here the ordering is between itinerant rather than localized electrons. NFL behavior has been established for $x$ in the neighborhood of 0.5.\cite{GTHM97} Recent measurements of the electrical resistivity and magnetic susceptibility of Ce(Ru$_{0.5}$Rh$_{0.5}$)$_2$Si$_2$ below 1~K\cite{TGOT01,TTM01} have been interpreted in terms of a quantum Griffiths-phase NFL mechanism\cite{CNJ00} for magnetic fields $\lesssim$1 T and quantum spin glass behavior at higher fields. 

At ${\rm temperatures} \gtrsim 2$~K $^{29}$Si NMR measurements in an aligned powder sample of Ce(Ru$_{0.5}$Rh$_{0.5}$)$_2$Si$_2$\cite{LMCN00} have shown that the local susceptibility is inhomogeneously distributed. The width~$\delta\chi(T)$ of this susceptibility distribution was found to be in good agreement with both the Griffiths-phase\cite{CNJ00} and so-called ``Kondo disorder''\cite{BMLA95,MDK97a} models of disorder-driven NFL behavior, which predict essentially the same $\delta\chi(T)$. In these theories a characteristic energy~$\Delta$ is inhomogeneously distributed in the sample. In the Griffiths-phase theory $\Delta$ is the tunneling energy~$E_t$ associated with a spin cluster, whereas in the Kondo-disorder model $\Delta$ is the Kondo temperature~$T_K$ of an individual spin. Each model yields a distribution function~$P(\Delta)$ that can be used to calculate sample averages of experimental quantities. For example, the sample-average ``bulk'' susceptibility~$\overline{\chi}(T)$ is given in terms of the local magnetic susceptibility~$\chi(T;\Delta)$ by
\begin{equation}
\overline{\chi}(T) = \int d\Delta\, P(\Delta)\chi(T; \Delta) \,.
\label{eq:chibar}
\end{equation}
For simplicity $\chi(T;\Delta)$ is often taken to be of the Curie-Weiss form
\begin{equation}
\chi(T; \Delta) = \frac{N(p_{\rm eff}\mu_B)^2}{3(T + \Delta)} \,,
\label{eq:curieweiss}
\end{equation}
where $p_{\rm eff}$ is the {\em f\/}-ion effective moment number.

In Ref.~\onlinecite{LMCN00} the data were analyzed under the assumption that the disordered susceptibility inhomogeneity is correlated only over short distances [``short range correlation'' (SRC), or static susceptibility correlation length~$\xi_\chi \lesssim {\rm lattice\ constant}\ a$]. The sensitivity of the NMR spectral width to $\xi_\chi$ comes about because a given $^{29}$Si nucleus is coupled to a limited number of neighboring Ce spins. If the susceptibilities of these Ce neighbors are uncorrelated or only slightly correlated because $\xi_\chi$ is short, then the interactions are averaged somewhat. If on the other hand $\xi_\chi$ is much longer than the distance to the Ce neighbors [``long range correlation'' (LRC), $\xi_\chi \gg a$], then the local coupling is not averaged and the (fractional) width of the frequency shift distribution is the same as that of the susceptibility distribution.\cite{LMCN00,MBL96} 

Agreement with disorder-driven models was obtained in the NMR experiments under the assumption of SRC\@. It was further assumed that the coupling could be characterized by an effective number~$n_{\rm eff}$ of Ce ions coupled with a fixed interaction strength to a given $^{29}$Si nucleus. For best agreement $n_{\rm eff} \approx 5$, which is reasonable crystallographically. It has been shown, however, that spectra from a second NMR nucleus or $\mu$SR can be used to test the SRC assumption.\cite{LMCN00,MBL96,MBAF96} 

This article reports measurements of magnetic susceptibility and $\mu$SR spectra for temperatures greater than $\sim$2~K in a high-quality single crystal of Ce(Ru$_{0.5}$Rh$_{0.5}$)$_2$Si$_2$, which complement the previous NMR measurements.\cite{LMCN00} Consistent with the NMR study, we find that the mean and width of the susceptibility distribution are in agreement with disorder-driven mechanisms. The predicted behavior is ``nearly NFL'' rather than NFL, i.e., the local susceptibility is Curie-Weiss-like with a distributed local Curie-Weiss temperature~$\Delta({\bf r})/k_B$, but the distribution function~$P(\Delta)$ vanishes at $\Delta = 0$. Then at low temperatures the system should revert to Fermi-liquid behavior in the absence of other mechanisms. 

The results of our susceptibility and transverse-field $\mu$SR (TF-$\mu$SR) experiments may be summarized as follows:

(1) The bulk susceptibility data can be fit to the Kondo-disorder and Griffiths-phase disorder-driven NFL models, thereby yielding the parameters of the distribution functions~$P(\Delta)$ that characterize these models. These distributions are narrower than found in our previous NMR investigations\cite{LMCN00} and possess little low-energy weight; the Griffiths-phase distribution, in particular, yields zero weight at $\Delta = 0$ rather than a Griffiths-McCoy singularity.\cite{CNJ00}

(2) The $\mu$SR linewidths agree with these distributions in the LRC limit, indicating a macroscopic distance scale to the susceptibility inhomogeneity responsible for the spread in muon Larmor frequencies. The SRC-limit inhomogeneity found in the NMR measurements apparently reflects differences in sample preparation. These differences preclude the independent test of the range of correlation (SRC vs LRC) discussed above. The amount of disorder derived from the disorder-driven models explains the nearly-NFL susceptibility of Ce(Ru$_{0.5}$Rh$_{0.5}$)$_2$Si$_2$ above $\sim$2~K, but is not capable of accounting for the NFL behavior found at low temperatures and fields.\cite{TGOT01,TTM01}

\section{Results}
\subsection{Static Susceptibility} \label{sect:suscept}
\begin{figure}[t]
\begin{center}
\epsfxsize 2.95in \leavevmode
\epsfbox{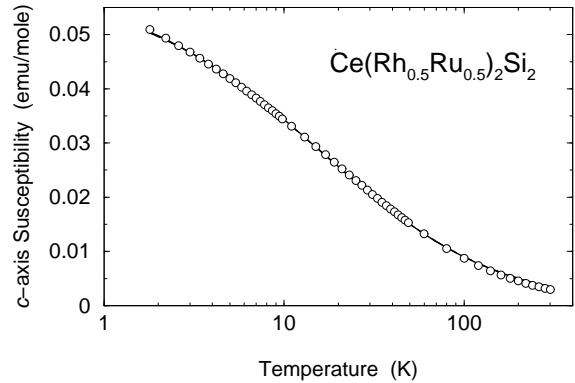}
\end{center}
\caption{Temperature dependence of $c$-axis bulk (sample-average) magnetic susceptibility~$\overline{\chi}_c$ in single-crystal Ce(Ru$_{0.5}$Rh$_{0.5}$)$_2$Si$_2$. Curve: fits to Griffiths-phase and Kondo-disorder models (indistinguishable on this plot).}
\label{fig:chi_vs_T}
\end{figure}

The temperature dependence of the bulk magnetic susceptibility~$\overline{\chi}_c$ in a field of 0.1 T applied parallel to the $c$ axis is given in Fig.~\ref{fig:chi_vs_T}.
These data were fit to the sample-average susceptibility~$\overline{\chi}(T)$ given by Eq.~(\ref{eq:chibar}) with distribution function~$P(\Delta)$ from both the Griffiths-phase and Kondo-disorder models.\cite{CNJ00,LMCN00,MDK97a} The Griffiths-phase picture ($\Delta = E_t$) yields\cite{CNJ00}
\begin{equation}
P(E_t) \propto \left\{ \begin{array}{ll}
E_t^{-1+\lambda} \,, & E_t < \omega_0 \,, \\
0 \,, & E_t > \omega_0 \,,
\end{array} \right.
\label{eq:grifdist}
\end{equation}
where $\omega_0$ is a high-frequency cutoff, so that $P(E_t)$ diverges as $E_t \to 0$ for values of the nonuniversal exponent~$\lambda < 1$ (Griffiths-McCoy singularities). In the Kondo-disorder model the Zener exchange coupling constant~$g = \rho{\cal J}$, where $\rho$ is the density of conduction-electron states at the Fermi energy and $\cal J$ is the conduction-electron--local-moment exchange interaction, is given a modest Gaussian distribution around a small average value.\cite{BMLA95,LMCN00} The local Kondo temperature
\[ T_K = E_F \exp(-1/|g|) \]
can then be widely distributed because of its exponential dependence on $g$. 

In both models the local susceptibility~$\chi(T;\Delta)$ is taken to have the Curie-Weiss form of Eq.~(\ref{eq:curieweiss}). 
A fit of Eq.~(\ref{eq:chibar}) to the bulk susceptibility then determines the parameters of the distribution function, which can be used to calculate the average of any function of $\Delta$ over the distribution. The fits to the Griffiths-phase and Kondo disorder models are given by the curve in Fig.~\ref{fig:chi_vs_T}, and are indistinguishable on this plot. 

\begin{figure}[t]
\begin{center}
\epsfxsize 2.95in \leavevmode
\epsfbox{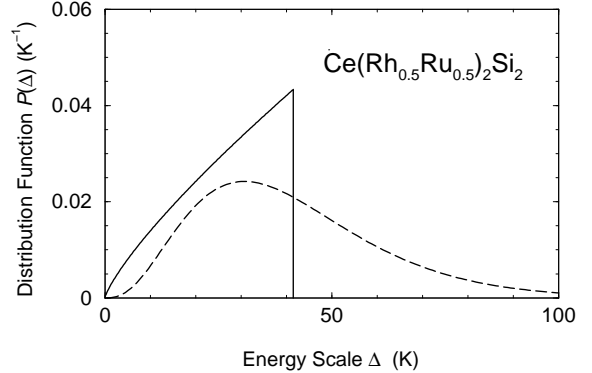}
\end{center}
\caption{Distribution functions~$P(\Delta)$ of characteric energies~$\Delta$ in 
single-crystal Ce(Ru$_{0.5}$Rh$_{0.5}$)$_2$Si$_2$, obtained from fits of Griffiths Phase (Ref.\protect\onlinecite{CNJ00}, solid curve) and ``Kondo disorder'' (Refs.\protect\onlinecite{BMLA95} and\protect\onlinecite{MDK97a}, dashed curve) disorder-driven NFL theories to bulk susceptibility data of Fig.~\protect\ref{fig:chi_vs_T}.}
\label{fig:distfunc}
\end{figure}

The fit parameters so obtained ($\lambda$, cutoff energy~$\omega_0$, and $p_{\rm eff}$ for the Griffiths-phase fit; average~$\overline{g}$ and standard deviation~$\delta g$ of $g$, $E_F$, and $p_{\rm eff}$ for the Kondo-disorder fit) are given in Table~\ref{tbl:fitparams} together with the corresponding values from fits to the susceptibility of the NMR sample (Ref.~\onlinecite{LMCN00}).
\begin{table}
\begin{tabular}{lccc}
& Parameter & $\mu$SR, & NMR, \\
& & single crystal\tablenote{Data of Fig.~\protect\ref{fig:chi_vs_T}.} & aligned powder\tablenote{Data of Ref.~\protect\onlinecite{LMCN00}.}\\
\noalign{\hrule}
Grifffiths & $\lambda$ & $1.8\pm0.2$ & 0.88 \\
Phase & $\omega_0$ (K) & $41\pm5$ & 170 \\
& $p_{\rm eff}$ & $3.01\pm0.06$ \\
\noalign{\hrule}
Kondo & $\overline{g}$ & $0.15\pm0.003$ & 0.16 \\
Disorder & $\delta g$ & $0.017\pm0.002$ & 0.021 \\
& $E_F$ (eV) & 1 (fixed) \\
& $p_{\rm eff}$ & $3.04\pm0.07$ \\
\end{tabular}
\vspace*{20pt}\caption{Parameters obtained from fits of Griffiths-phase and Kondo Disorder disorder-driven NFL models to bulk susceptibility data. See text for definitions.}
\label{tbl:fitparams}
\end{table}
The distribution functions~$P(\Delta)$ from the present results are given in Fig.~\ref{fig:distfunc}. 

The most important feature of these results is that for both models $P(\Delta)$ has low weight for small $\Delta$. For Kondo-disorder fits this result is in qualitative agreement with the NMR-sample data,\cite{LMCN00} which also yielded narrow distributions. But the best Griffiths-phase fit is obtained for $\lambda > 1$, i.e., for a zero rather than a singularity at $P(E_t{=}0)$ [cf.\ Eq.~(\ref{eq:grifdist})], whereas the NMR-sample susceptibility was best fit with a singular form of  $P(E_t)$ ($\lambda = 0.88$). This indicates that within the framework of these disorder-driven models for NFL behavior the $\mu$SR sample exhibits only ``nearly-NFL'' behavior, since nonzero values of $P(\Delta{=}0)$ are necessary if the lowest-lying excitations are to be local-moment-like or cluster-like rather than Fermi-liquid in character. The theoretical distributions are simply too narrow to yield the NFL behavior observed at low temperatures.\cite{TGOT01,TTM01}

\subsection{\boldmath NMR and $\mu$SR Frequency Shifts} \label{sect:shifts}
In a paramagnetic sample spin probe (nucleus or muon) frequency shifts reflect the electronic spin polarization to which the spin probes are coupled via a combination of dipolar and indirect hyperfine interactions.\cite{Sche85} A given spin probe (nucleus or muon)~$i$ is coupled to neighboring {\em f\/}-electron moments~$j$, resulting in a linear relation between the spin-probe frequency shift~$K_i$ and the {\em f\/}-electron susceptibility~$\chi_j$:\cite{LMCN00,MBL96,MBAF96}
\[ K_i = \sum_j a_{ij}\chi_j \,, \]
where the coupling constants~$a_{ij}$ can be expressed in terms of corresponding coupling fields~$H_{ij}^{\rm coup} = N\mu_Ba_{ij}$; $N$ is Avogadro's number if the $\chi_j$ are expressed in molar units. We can write $H_{ij}^{\rm coup}$ as the sum of transferred-hyperfine and dipolar contributions~$H_{ij}^{\rm thf}$ and $H_{ij}^{\rm dip}$, respectively.

As described in Sect.~\ref{sect:Intro} above and elsewhere,\cite{LMCN00,MBL96,MBAF96,BMAF96} the relation between an inhomogeneous distribution of $\chi_j$s and the resulting distribution of $K_i$s depends on the range of correlation of the inhomogeneous susceptibility. In the two extreme limits of SRC and LRC the rms spreads $\delta K$ and $\delta\chi$ are simply related:\cite{LMCN00,MBL96,MBAF96}
\[ \delta\chi = \delta K/a^{*} \,, \]
where the effective coupling constant~$a^{*}$ depends on the correlation length~$\xi_\chi$:
\begin{equation}
a^{*} = \left\{\begin{array}{ll}
a_{\rm LRC} = \left| \sum_j a_{ij} \right| & \mbox{(LRC)} \,, \\
a_{\rm SRC} = \left( \sum_j a_{ij}^2 \right)^{1/2} & \mbox{(SRC)} \,.
\end{array} \right.
\label{eq:astar}
\end{equation}
In the LRC limit this yields 
\begin{equation}
(\delta\chi/\overline{\chi})_{\rm LRC} = \delta K/|\overline{K}| \,, 
\label{eq:LRC}
\end{equation}
and in the SRC limit we can write 
\begin{equation}
\delta\chi/\overline{\chi}_{\rm SRC} = (a_{\rm LRC}/a_{\rm SRC})(\delta K/|\overline{K}|) \,.
\label{eq:SRC}
\end{equation}
We will use these relations in the analysis of our $\mu$SR results.

\subsection{\boldmath Transverse-Field $\mu$SR Spectra}
The $\mu$SR experiments were carried out at the M20 muon channel at TRIUMF, Vancouver, Canada. A magnetic field~${\bf H}_0$ was applied parallel to the $c$ axis of the single-crystal sample, and the muon spin~${\bf S}_\mu$ was oriented perpendicular to ${\bf H}_0$. In this TF-$\mu$SR configuration the width of the $\mu$SR resonance reflects the distribution of muon frequency shifts in the sample as long as lifetime broadening (spin-lattice relaxation) is negligible, which has been confirmed by zero-field $\mu$SR measurements.\cite{YMTK99} 
 
\begin{figure}[t]
\begin{center}
\epsfxsize 2.95in \leavevmode
\epsfbox{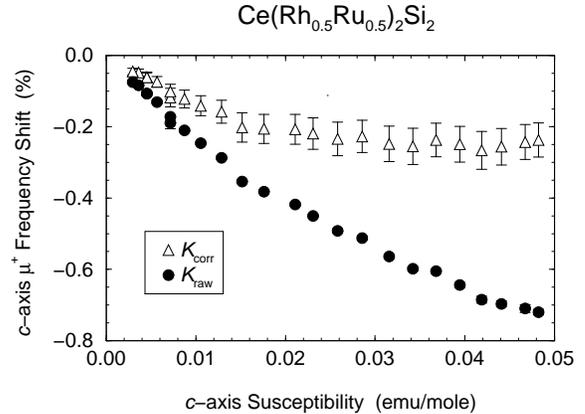}
\end{center}
\caption{Dependence of sample-average muon $c$-axis frequency shift $\overline{K}_c$ on $c$-axis bulk susceptibility~$\overline{\chi}_c$ in Ce(Ru$_{0.5}$Rh$_{0.5}$)$_2$Si$_2$, with temperature an implicit parameter. Circles: raw data. Triangles: data corrected for Lorentz and demagnetizing fields. The error bars include uncertainty in the value of the demagnetization coefficient.}
\label{fig:Kc_vs_chi}
\end{figure}

The spectra were fit to both Lorentzian and Gaussian lineshapes; values of the goodness-of-fit parameter~$\chi^2$ were not appreciably different for these choices.  Figure~\ref{fig:Kc_vs_chi} gives the dependence of the average (line centroid) fractional frequency shift $\overline{K}_c$ on the bulk $c$-axis susceptibility~$\overline{\chi}_c$ in Ce(Ru$_{0.5}$Rh$_{0.5}$)$_2$Si$_2$, with temperature an implicit parameter, before and after correction for Lorentz and demagnetizing fields.
The curvature of the corrected $\overline{K}_c$ vs $\overline{\chi}_c$ could be due to a number of causes, including crystal-field effects and a Curie ``tail'' in the susceptibility from trace metallurgical phases.\cite{LMCN00} Above $\sim$20~K $\overline{K}_c$ is approximately proportional to $\overline{\chi}_c$, with a slope that yields a coupling field
\begin{eqnarray}
H_c^{\rm coup} & = & N\mu_B \sum_j a_{ij} \label{eq:coupfield} \\
& = & -0.71\pm0.06\ {\rm kOe}/\mu_B \,. \nonumber
\end{eqnarray}
This is considerably more negative than the value $-0.38\ {\rm kOe}/\mu_B$ found in undoped CeRu$_2$Si$_2$.\cite{Amat97} and indicates that $H_c^{\rm thf}$ is weaker in Ce(Ru$_{0.5}$Rh$_{0.5}$)$_2$Si$_2$. This may reflect a difference in Ce--Si hybridization or, alternatively, a different muon stopping site in Ce(Ru$_{0.5}$Rh$_{0.5}$)$_2$Si$_2$.\cite{Heff00} Indeed, different muon stopping sites are found in the end compounds~CeRu$_2$Si$_2$ ($\langle\frac{1}{2}\, \frac{1}{2}\, 0\rangle$, Wyckoff notation $2b$)\cite{Amat97} and CeRh$_2$Si$_2$ (two sites: $\langle\frac{1}{2}\, 0\, 0\rangle$, $4c$, and $\langle\frac{1}{2}\, \frac{1}{4}\, 0\rangle$, $4d$).\cite{DdRSYC90}

\begin{figure}[t]
\begin{center}
\epsfxsize 2.95in \leavevmode
\epsfbox{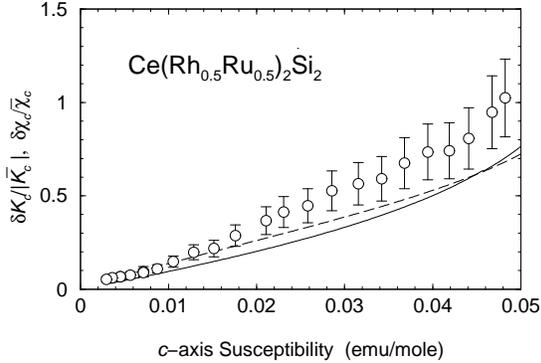}
\end{center}
\caption{Data points: dependence of estimator~$\delta K_c/|\overline{K}_c|$ of fractional $c$-axis susceptibility spread~$\delta\chi_c/\overline{\chi}_c$ on $c$-axis bulk susceptibility~$\overline{\chi}_c$, with temperature an implicit parameter, in Ce(Ru$_{0.5}$Rh$_{0.5}$)$_2$Si$_2$. Curves: dependence of $\delta\chi/\overline{\chi}$ on $\chi$ from fits of Griffiths-phase (solid curve) and Kondo-disorder (dashed curve) models to bulk susceptibility.}
\label{fig:dKonK_vs_chi}
\end{figure}

The ratio of the fractional muon linewidth~$\delta K_c$ to the frequency shift magnitude~$|\overline{K}_c|$ is plotted vs $\overline{\chi}_c$ in Fig.~\ref{fig:dKonK_vs_chi}. The increase of $\delta K_c/|\overline{K}_c|$ with increasing susceptibility is a salient feature of disorder-driven models for NFL behavior.\cite{BMLA95,MBL96}  The curves in Fig.~\ref{fig:dKonK_vs_chi} give the fractional rms width~$\delta\chi/\overline{\chi}$ of the susceptibility distribution from the disorder-driven NFL distributions discussed in Sect.~\ref{sect:suscept}, where $\delta\chi$ is obtained from the relation
\[\delta\chi(T) = \left\{ \int d\Delta\, P(\Delta) \left[ \chi(T; \Delta) - \overline{\chi}(T) \right]^2 \right\}^{1/2}
\]
[cf.\ Eq.~(\ref{eq:chibar})].

The data are in reasonable agreement with the disorder-driven pictures in the LRC limit, for which $\delta K/|\overline{K}|$ (data points in Fig.~\ref{fig:dKonK_vs_chi}) is the estimator of $\delta\chi/\overline{\chi} [Eq.~(\ref{eq:LRC})]$. In the SRC limit this estimator is obtained by multiplying the experimental values of $\delta K/|\overline{K}|$ by $a_{\rm LRC}/a_{\rm SRC}$ [Eq.~(\ref{eq:SRC})]. The magnitude of the slope of $\overline{K}_c$ vs $\overline{\chi}_c$ gives the value of $a_{\rm LRC}$.\cite{LMCN00,MBL96,MBAF96} It is more difficult to obtain $a_{\rm SRC}$, but we can estimate separately the dipolar and transferred hyperfine contributions to this quantity and hence to $a_{\rm LRC}/a_{\rm SRC}$. For dipolar coupling the value of this factor obtained from lattice sums [Eq.~(\ref{eq:astar})] is 1.51. For a constant transferred hyperfine coupling strength to $n_{\rm eff}$ near neighbors the factor is $\sqrt{n_{\rm eff}}= 2$--2.2 for the crystallographically reasonable range~$n_{\rm eff} = 4$--5.\cite{LMCN00} Thus in the SRC limit the experimental estimator of $\delta\chi/\chi$ is significantly increased, and agreement with the theoretical curves is worsened, for both dipolar and transferred hyperfine coupling. This is evidence that in the present single-crystal sample the LRC limit is appropriate, i.e., that there is little effect of atomic-scale disorder on the local static susceptibility.

\section{Conclusions}
The fractional width~$\delta\chi_c/\overline{\chi}_c$ of the inhomogeneous susceptibility distribution in Ce(Ru$_{0.5}$Rh$_{0.5}$)$_2$Si$_2$ is considerable ($\sim$100\% at low temperatures, see Fig.~\ref{fig:dKonK_vs_chi}), but the parameters of the distribution functions (Table~\ref{tbl:fitparams}) indicate smaller widths (smaller $\omega_0$ and larger $\lambda > 1$ for the Griffiths-phase fit; smaller $\delta g/\overline{g}$ for the Kondo-disorder fit) in the $\mu$SR single crystal than in the NMR aligned powder sample of Ref.~\onlinecite{LMCN00}. The origin of the disorder may also be different in the two samples, since the fits agree better with the LRC limit in the present work but with the SRC limit in the NMR study. This suggests that local disorder in the Ce-moment--conduction-electron hybridization strength is significant in the NMR sample but relatively weak in the $\mu$SR single crystal, although variations over length scales longer than a few lattice parameters remain as a source of inhomogeneity in the latter sample.

In both samples the low probability of small energy scales (cf.\ Fig.~\ref{fig:distfunc}) is suggestive of low spectral densities of spin fluctuations, quantum or thermal, at low frequencies. This is in agreement with the relatively slow $\mu$SR spin-lattice relaxation rates observed in Ce(Ru$_{0.5}$Rh$_{0.5}$)$_2$Si$_2$,\cite{YMTK99} since the muon relaxation rate is proportional to the strength of the thermal noise spectrum at the low muon Larmor frequency.\cite{Sche85} 

The data taken as a whole indicate that the disordered susceptibility revealed by the magnetic resonance ($\mu$SR and NMR) experiments, while significant, is not likely to be related to the NFL behavior observed in Ce(Ru$_{1-x}$Rh$_x$)$_2$Si$_2$ at temperatures below 1~K and magnetic fields below 1 kOe.\cite{TGOT01,TTM01} The latter exists over a range of Rh concentration~$x = 0.5$--0.6, as expected from the Griffiths-phase picture,\cite{TTM01} but the large clusters with low characteristic energies needed in this scenario are simply not observed in magnetic resonance experiments at higher temperatures and fields. Furthermore, it is unlikely that Griffiths-phase clusters form only for $T \lesssim 1$~K, since this temperature is considerably smaller than the energy scale for the disorder ($\sim$30~K, cf.\ Fig.~\ref{fig:distfunc}). Magnetic-resonance shift and linewidth measurements would be desirable as direct probes of disorder in the low-temperature low-field NFL region, but unfortunately accurate data cannot be obtained at the low fields ($\sim$100 Oe) required for the NFL behavior.

\medskip We are grateful to B. Hitti, M. Good, and S. R. Kreitzman, TRIUMF, for help with the $\mu$SR experiments. This research was supported in part by the U.S. NSF, Grants DMR-9731361 (U.C. Riverside) and DMR-9820631 (CSU Los Angeles), and by a Grant-in-Aid for Scientific Research from the Japanese Ministry of Education, Science and Culture, and was performed in part under the auspices of the U.S. DOE.

\end{document}